\documentclass[12pt,preprint]{aastex}

\begin{document}

 \title{Image Slicer Performances from a Demonstrator for the SNAP/JDEM Mission \\
    Part I: Wavelength Accuracy }

\author{M-H. Aumeunier\altaffilmark{1,3},  A. Ealet\altaffilmark{2}, E. Prieto\altaffilmark{1}, C.Cerna\altaffilmark{2}, P-E. Crouzet\altaffilmark{2} }

\altaffiltext{1}{Laboratoire d'Astrophysique de Marseille, 38, rue Fr\'ed\'eric Joliot-Curie, F-13388 Marseille cedex 13, France}
\altaffiltext{2}{Centre des Physiques des Particules de Marseille, 163 av. de Luminy, Case 902, 13288 Marseille cedex 09, France}
\altaffiltext{3}{marie-helene.aumeunier@oamp.fr}

\begin{abstract}
A well-adapted visible and infrared spectrograph has been developed for the SNAP (SuperNova/Acceleration Probe) experiment proposed for JDEM. The instrument should have a high sensitivity to see faint supernovae but also a good redshift determination better than 0.003(1+z) and a precise spectrophotometry (2\%). An instrument based on an integral field method with the powerful concept of imager slicing has been designed. A large prototyping effort has been performed in France which validates the concept. In particular a demonstrator reproducing the full optical configuration has been built and tested to prove the optical performances both in the visible and in the near infrared range. This paper is the first of two papers. The present paper focus on the wavelength measurement while the second one will present the spectrophotometric performances. We adress here the spectral accuracy expected both in the visible and in the near infrared range in such configuration and we demonstrate, in particular, that the image slicer enhances the instrumental performances in the spectral measurement precision by removing the slit effect.
This work is supported in France by CNRS/INSU/IN2P3 and by the French spatial agency (CNES) and in US by the University of California. 

\end{abstract}

\keywords{Integral Field Spectrograph, slicer, spectro-photometric calibration, SNAP, supernovae, wavelength calibration}

\section{Introduction}

Integral Field Spectrograph (IFS) is a powerful technique able to provide simultaneous the spectrum of each spatial sampling element in a given field of view (FoV). The principle consists in rearranging the 2D spatial field thanks to a Integral Field Unit and then in using a classical spectrograph to disperse the light. It exits three types of Integral Field Unit using microlenses array, fibers combined with microlenses and image slicer.The use of image slicer is very suitable to measure faint object thanks to its high efficiency and to minimize the constraint on the pointing accuracy.
Since about ten years, two technologies of image slicer have been developed simultaneous. The first one uses a slicing-mirror
stack in aluminum (``monolithic design'') using diamond turning techniques. The second one uses individual mirrors on ZERODUR assembled using optical-bonding techniques (molecular adhesion). Early prototypes  have shown the capability of the image slicer in ground-based astronomy for the Infrared measurement (see the instruments SPIFFI \cite{Eisenhauer2000} or NIFS \cite{hart2003}. The aluminum monolithic approach is also planned to be used for MIRI  in the Mid-Infrared \cite{wells2006_spie} and for NIRSpec in the Near-Infrared \cite{prieto2006_OE} of the future James Webb Space Telescope (JWST). However, the aluminum has poor surface roughness quality under $1.5 \; \mu m$ which limits the use of this technology. The ZERODUR approach has also keep  on maturing and has been developed in the visible range. A first prototype has been build and tested in the frame of European Spatial Agency (ESA) for the Near-InfraRed Spectrometer(NIRSpec) of the JWST and has demonstrated the feasibility of an imager slicer for space application  \cite{Laurent2004_spie_nirspec}. After further prototyping, this technology has been selected for the Multi Unit Spectroscopic Explorer (MUSE) for the second-generation VLT \cite{laurent2006_paps} working in the visible. In the mean time, a complete IFS-slicer demonstrator of spectrograph has been designed, manufactured and tested in the visible and infrared range for the SuperNoave/Acceleration Probe(SNAP). In this paper, we present some results of this demonstrator both in the visible and infrared range.

The SNAP satellite is designed to measure very precisely the cosmological parameters and to determine the nature of the dark energy [\textit{http://snap.lbl.gov/}]. The mission includes the measurement of some 2000 supernovae (SNe) of Type Ia up to a redshift of z=1.7. Details of the mission and the expected physics results can be found \cite{aldering2004}. Spectroscopy of each candidate supernova near maximum light is required to identify and control intrinsic variations through spectral features. The spectrograph is also required to measure the redshift of the host galaxy at a precision better than $0.003 \times \left(1+z\right)$. The optical properties include high efficiency, very low noise and accurate spectro-photometric calibration at some percent level. The slicer has shown to have many advantages compared to a standard slit configuration but this technology should be validated and its performances should be tested for space application. A large prototyping effort is going on in France to bring this technology at this level \cite{pamplona2008_spie}. In this paper, we focus on the slicer optical properties in a full spectrograph configuration.

The SNAP project has designed a specific demonstrator, reproducing the spectrograph concept \cite{prieto2008_spie} and including a new slicer unit.  An optical bench using a new slicer prototype has been manufactured to evaluate the optical performances and to test the accuracy of the spectro-photometric calibration. We present in this paper the first results for wavelength determination both in the visible and in the IR.

After recalling the spectrograph requirement concept and design, we present the demonstrator set up and the tools developed to test the optical performances. We describe a full simulation which includes the optical and detector effects, developed to  test performances and adapted to the demonstrator to prepare procedures. The simulation  has been compared later with data . We show in particular that the optical characteristics and performances of the demonstrator are well reproduced, which proves that we control well the optical properties of the instrument. 

We present then the first results in the visible and infrared focusing on the wavelength determination. We describe the wavelength calibration procedure, adapted to our configuration. We test then the accuracy on the wavelength of a point like source. We show there than the wavelength can be recover at better than 1 nm, on the full range thanks to the slicer properties. In particular we demonstrate than the slicer provides a natural spatial and spectral dithering that correct the classical slit effect without any mechanism.

\section{Spectrograph description}

The instrument should be well adapted to space environment (small, compact, light). To see faint supernovae and galaxies, a low spectral resolution  covering the visible and the near infrared  range, with very high optical and detector performances (the main limitation is the telescope diameter)  and a constant resolving power in the 0.6-1.7 $\mu m$ range. Each supernova and its host galaxy should be exposed simultaneously to minimize exposure time. The spectrograph is also a key component of the calibration procedure. The spectrograph will be used to transfer the calibration of fundamental standard stars to primary standards in the range $M_V= 12-18$ mag where the imager cannot reach the needed sensitivity. This requires a spectro-photometric calibration at $\approx$ 2\% accuracy with a wavelength calibration at 1/10 of a pixel (about 1 nm).

Given the science drivers and specifications, we have conducted a trade-off study to choose the best instrument concept. The requirement for simultaneous acquisition of SN and host spectra, and for the level of needed precision lead us to prefer a 3D spectrograph to a traditional long slit spectrograph. The specification are summarized on the table \ref{tab:param_spectro}. 

\begin{figure}[htbp]
	\centering
	\includegraphics[angle=0,width=0.45\textwidth]{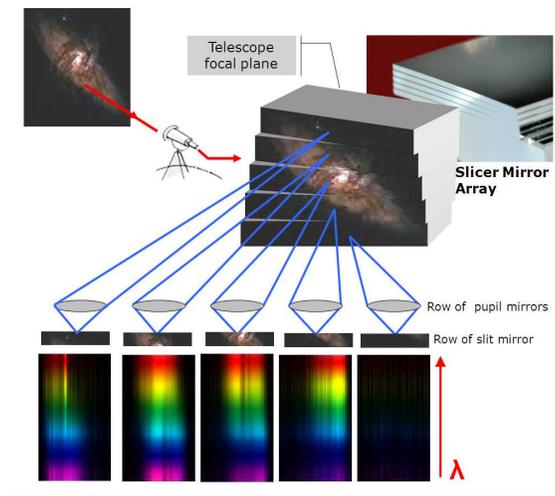} 
	\caption{Slicer principle: the FoV is sliced into $N$ strips on a slicing mirror, creating $N$ telescope pupil images in the pupil plane. Because of a tilt adapted to each slice, the images are arranged along a line and form a pseudo-slit.}
	\label{fig:slicer_principle}
\end{figure}

A 3D spectrograph reconstructs the data cube including the two spatial directions X and Y plus the wavelength direction. For each spatial pixel, the spectrum is reconstructed.  Thanks to the large field of view ($3\times6$ arc seconds), the pointing requirements are relaxed. The image slicers minimize optical losses and improve the efficiency and the compactness of the system. Figure \ref{fig:slicer_principle} shows the principle of this technique. The field of view is sliced along N (for SNAP N=40) strips on a ``slicing mirror". Each of N slices re-images the telescope pupil, creating N telescope pupil images in the pupil plane. Thanks to a tilt adapted to each individual slice, the N pupil images lie along a line. These images are arranged along a line and form a ``pseudo-slit".
 
The spectrograph images then the entrance slit on the detector after a passage through a dispersing prism, for constant and low resolution (R$\approx$100) and high optical efficiency. The spatial and spectral resolution are optimized to maximize the signal-to-noise ratio for the faint object (SNIa up to 26 magnitude at $z=1.7$). The spatial resolution is of 0.15 ${\rm arc second}^2$. The slit is then imaged on 2 pixels in the visible and only on one in the IR to enhance signal to noise performance. Each slit covers  then 20 pixels in the spatial direction and 1(2) pixel for IR (visible) respectively in the dispersion direction. The spectrograph illuminates then two HgCdTe detectors from Teledyn working from 1 to 1.7 $\mu m$ and two visible CCD detectors from LBL covering 0.4 to 1 $\mu m$. The total size including detectors is of 230x250x100 mm for a weight of 10 kg. 
 
 The PSF\footnote{Point Spread Function} is optimized to be diffracted limited at $1 \mu$m, then the spectral PSF, which is cut by the slit, have a size smaller than the slit width when $\lambda \le 1 \; \mu$m and is sampled on only one pixel in the infrared range, under the Nyquist criteria (in this paper, we will call it an 'under sampled configuration'). Then, the wavelength measurement will be more difficult than in standard configuration. This implies the point-like source will be strongly sensitive to the slit effect both in the visible (see paragraph \ref{sec:slit_effect}) because the PSF is small and in the infrared because the slit width is imaged on only one pixel. 

\section{The demonstrator}

In order to validate \textit{in-situ} the instrument performances, we have build a complete prototype of an integral field spectrograph in the visible and infrared range \cite{Aumeunier2006_spie}. The demonstrator characteristics are the same ones as the on-board SNAP spectrograph:  low and constant spectral resolution and under-sampled PSF in the infrared range. Table \ref{tab:param_spectro} summarizes the main specifications of the SNAP spectrograph on-board and the one of the demonstrator. The two main changes are: the Field Of View (FoV) has been reduced to 5 slices for practical accommodation and the telescope aperture and the pointing system has been replaced by a steering mirror with a diaphragm. 
The diaphragm aperture is chosen to reproduce the telescope PSF size in the slicer plane. The steering mirror is equipped with a servo-control motor working on real time and can scan the complete FoV in the two directions with an accuracy better than 1/70 of a slice width. 

\begin{figure}[htbp]
	\centering
	\begin{tabular}{cc}
	\includegraphics[width=0.48\textwidth]{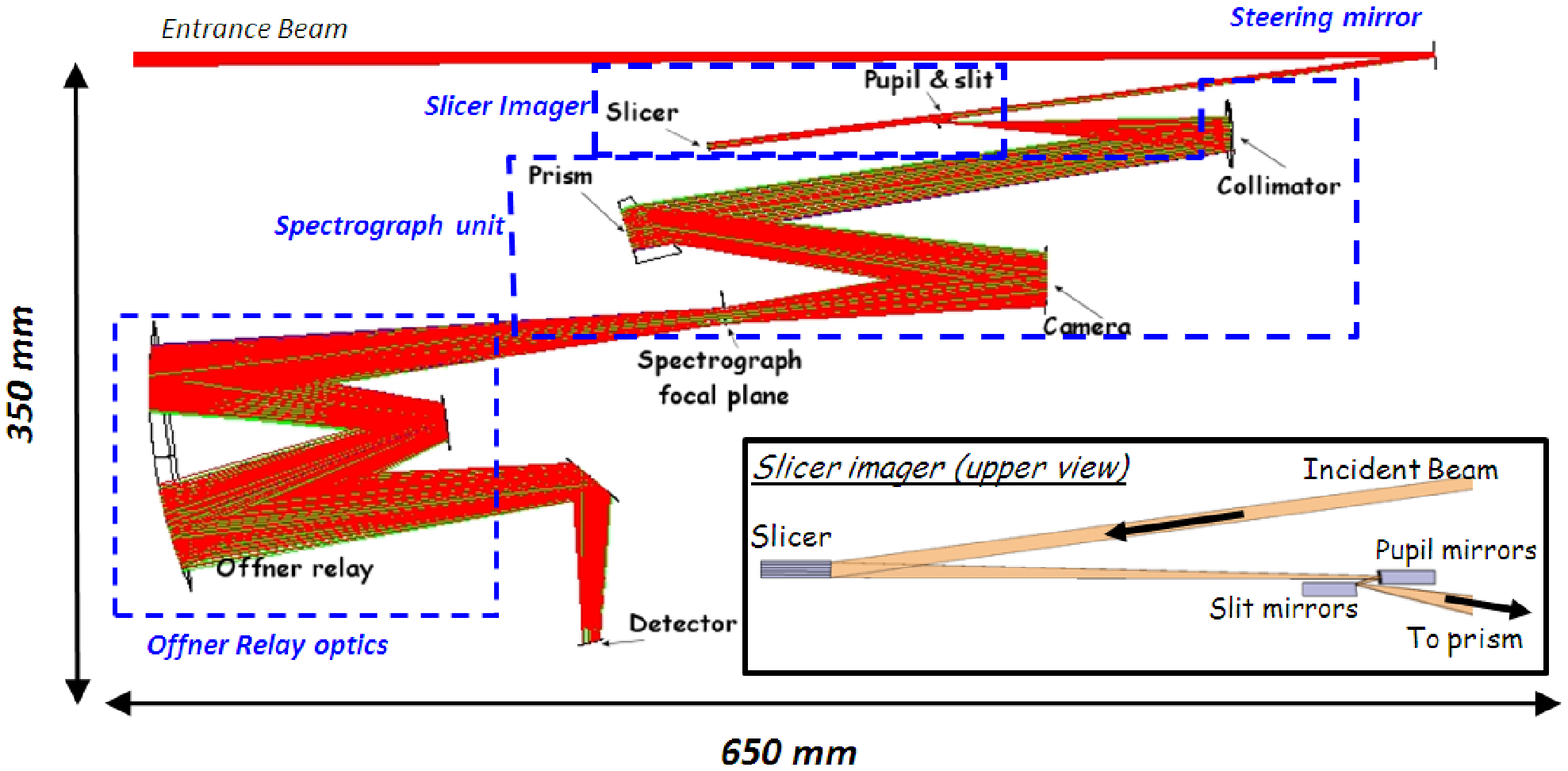} & \includegraphics[width=0.38\textwidth]{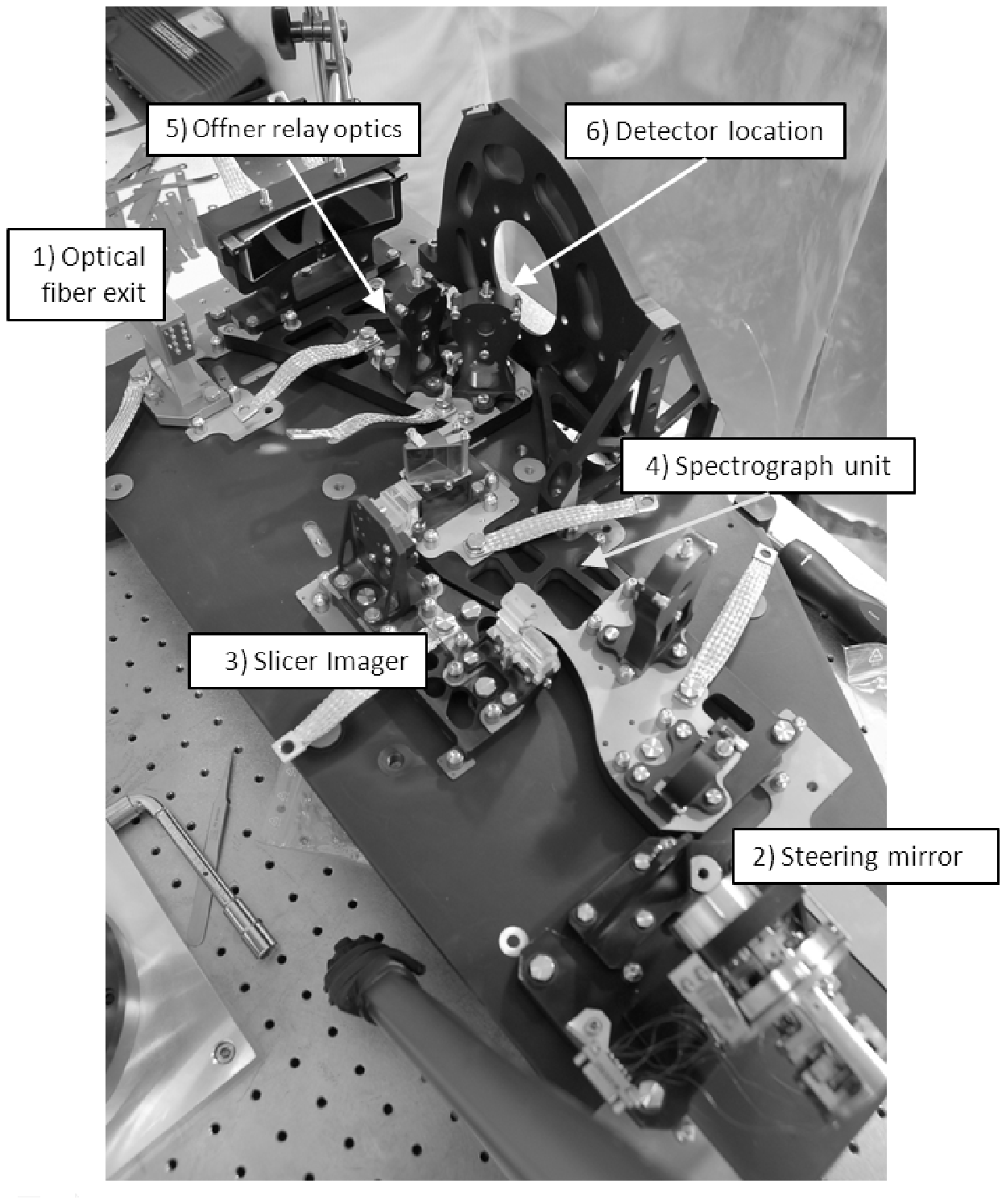} \\ 
		\end{tabular}
	\caption{Demonstrator. \textit{Left}: Optical design. \textit{Right}: Optical Bench picture.}
	\label{fig:demonstrator_picture}
\end{figure}

The input source is a QTH lamp combined with a tunable monochromator which  produce emission lines  of five nanometers width. The line positions are known  at 0.35 nm and the width at 2 nm. Finally, optical fibers placed at the monochromator exit allow to have collimated point-like sources. The details of the set-up are presented in \cite{cerna2008_spie}. The construction assembly has been done from 2006 to 2007. Fig.\ref{fig:demonstrator_picture} shows a picture of the complete demonstrator optical bench.

We have realized two campaigns of measurements. The first one has been done at room temperature using a CCD camera (Apogee Kodak), the second one in cryogenic environment in the infrared range using a last generation HgCdTe of IR detector 2k$\times$2k developed by Rockwell/Teledyne \cite{Barron2007}. A commissioning campaign has been done to test the functionalities. Results are presented in \cite{cerna2008_spie}. This confirms that the set-up was working well. We focus in this paper on the wavelength calibration procedure, test and results. 

 \begin{table*}[htbp]
\tiny
\begin{center}
\begin{tabular}{|l|l||c|c||c|c|}
\hline
 \multicolumn{2}{|c||}{} & \multicolumn{2}{c||}{\small SNAP spectrograph} & \multicolumn{2}{c|}{\small Demonstrator} \\
\cline{3-6}
 \multicolumn{2}{|c||}{} & \small Visible & \small Near Infrared & \small Visible & \small Near Infrared  \\
\hline
\small Field Of View  & \small Specification &  \multicolumn{2}{c||}{\small 3 $\times$ 6 $\rm{arcsec}^2$} & \multicolumn{2}{c|}{\small 3 $\times$ 0.75 $\rm{arcsec}^2$}\\
\hline
\small F-ratio  &  \small Specification  & \multicolumn{2}{c||}{\small 344} & \multicolumn{2}{c|}{\small 344} \\
 & \small Description  & \multicolumn{2}{c||}{\small telescope + relay optic}  &  \multicolumn{2}{c|}{\small specific illumination unit} \\
\hline
\small Spatial  & \small  Specification & \multicolumn{2}{c||}{\small 0.15 arcsec} & \multicolumn{2}{c|}{\small 0.15 arcsec}\\
 Resolution &  \small Description  & \multicolumn{2}{c||}{\small 2$\times$20 slices of 0.5$\times$10 mm size} &  \multicolumn{2}{c|}{\small 5 slices of 0.5$\times$10 mm size}\\
 \hline
\small Wavelength   & \small Specification   & \small [0.35-0.98 $\mu m$]   & \small [0.98-1.7 $\mu m$] &\small [0.35-0.98 $\mu m$]  & \small [0.98-1.7 $\mu m$] \\
\cline{3-6}
  coverage & \small Description  & \multicolumn{2}{c||}{\small two optical arms } &\multicolumn{2}{c|}{\small one optical arm} \\
 \hline
\small Spectral   & \small Specification  & \small 200   & \small 70   & \small 200   & \small 70\\
\cline{3-6}
 Resolution   & \small Description  &  \multicolumn{2}{c||}{\small Prism}  &  \multicolumn{2}{c|}{\small Prism} \\
 \hline
\small Pixel size & \small  Specification  &\small 9 $\mu m$ &  \small 18 $\mu m$  &  \small 9 $\mu m$  & \small  18 $\mu m$ \\
\cline{3-6}
    & \small Description & \multicolumn{2}{c||}{\small Camera F-ratio=12 } & \multicolumn{2}{c|}{\small Camera F-ratio=12 }\\
  \hline
\multicolumn{2}{|c||}{\small Detector temperature}  &  \multicolumn{2}{c||}{\small Passive Cooling}   & \small  Room temperature &  \small  140 K \\
\hline
\multicolumn{2}{|c||}{\small Throughput}  &  \multicolumn{2}{c||}{\small $\geq \; 40 \; \% $}   &  \multicolumn{2}{c||}{\small $\geq \; 40 \; \% $}    \\
\hline
\end{tabular}
\end{center}
\caption{Specifications for the SNAP spectrograph (on flight) and demonstrator  }
\label{tab:param_spectro}
\end{table*}

\section{Simulation}

We have developed a complete simulation of the instrument. This simulation takes both the optical and detector effects into account. This allows to verify our understanding of the optics. This was used first to prepare tests and procedures, and to  compare experimental data to validate the optical design and performances.

The optical simulation of the instrument response is based on the modeling of the optical response of a monochromatic point source. Its intensity, size and shape are driven by the optical effects such as diffraction, aberrations and distortions and varies with the spatial position within the Field Of View(FoV) and the wavelength (paragraph \ref{ss-sec:optical modeling}). Then, the PSF is sampled on the detector taking the pixel response into account (paragraph \ref{ss-sec:detector}). From these PSFs, we can reconstruct the images of spatial extended sources.

\subsection{PSF modeling}
\label{ss-sec:optical modeling}

To model the demonstrator PSF, we have used the optical ray-tracing model, Zemax. This is used classically to describe the light ray propagation according to the principle of the geometrical optics. We have used here a recent extension of Zemax, named POP (Physical Optics Propagation) to take into account  the diffraction phenomenon. POP is modeling the optical systems by propagating the wavefront. The beam is propagated through the free space between optical surfaces using the Fresnel diffraction propagation and, at each optical surface, a transfer function is computed which transfers the beam from one side of the optical surface to the other \cite{zemax2005}. It provides the beam amplitude and the phase shift for any optical plane of the optical system.
 
\begin{figure}[th]
	\centering
	\includegraphics[angle=0,width=0.5\textwidth]{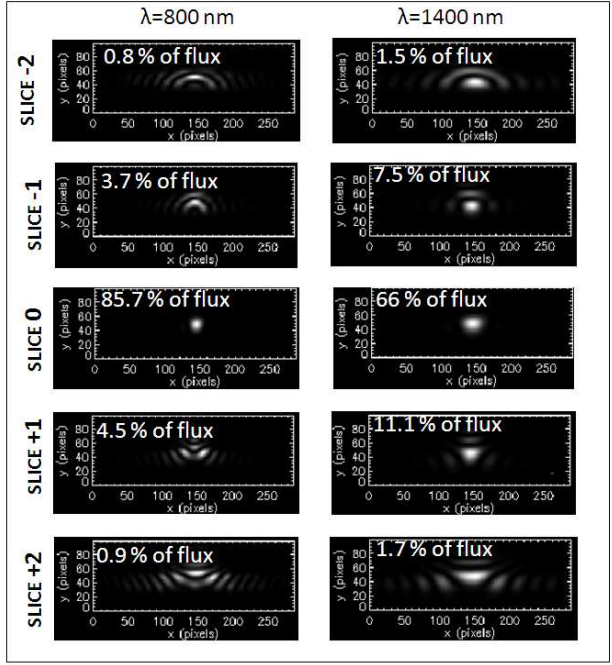} 
	\caption{ Examples of simulated PSFs (oversampled: one pixel= 1$\mu m$) on the exit focal plane of the SNAP spectrograph in the visible and infrared arm when the point-like source is at the slice center (number 0). The PSF is spread out on 5 slices. More 85\% of the incident flux is recovered taking into account than the 5 slices.}
	\label{fig:psfsnap}
\end{figure}

In the IFS-slicer, the telescope PSF is shared out on N ``slices'' and each part of the telescope PSF inside the slices is then convoluted with the spectrograph PSF. The PSF is a diffraction-limited PSF over all the wavelength range. Fig.\ref{fig:psfsnap} shows the simulated PSF in the exit focal plane of the spectrograph in the visible and infrared range. We look at the PSF spread out on five slices when the point source is located at the center of the central slice. At this point, the PSF is symmetric and the PSF core with more than 50 \% of intensity is imaged on only one slice, the nearest slices contain the PSF rings. 
The flux losses are in this way minimized thanks to the slicer. Contrary to a slit-spectrograph, almost all the incident flux is recovered by taking into account the nearest slices. The total optical throughput is greater than 90\% for the shortest wavelength and greater than 85\% for the longest wavelength.

\subsection{Discussion on the PSF modeling}
\label{ss_sec:psf_model}

The POP tool is very attractive because it takes into account both the diffraction and the aberrations without doing approximations. The main limitation is that the PSF modeling quality is very sensitive to the beam sampling and the support (diaphragm) size to translate to the Fourier space. Nevertheless, these are parameters difficult to control with POP and this can lead to aliasing phenomenon on the exit PSF. Practically, the user can choose the first support size and the input beam sampling, the POP tool computes then automatically the size and the sampling for the other surfaces. This may be more difficult to use for complex optical system.

In order to check the PSF computed with POP, we have also developed a complete simulation using the Fourier formalism with a computing software IDL\footnote{Interactive Data Language}. The complex amplitude is thus propagated through each optical sub-system (telescope, slicer unit, spectrograph) using the Fraunhofer approximation. This is allowed because each optical plane is located at the conjugated plane of the previous one. The advantage is that such a simulation allows a good control of the PSF amplitude through the instrument whereas it is not possible with POP. The disadvantages is that the aberrations are not directly taken into account. The aberrations are computed independently with Zemax by using the Zernike polynomials. They are then implemented in a phase map $W$ in the pupil function as $W\left( {x,y} \right) = \sum\limits_i {a_i \left( \lambda  \right)}  \times Z_i \left( {x,y} \right)$ where  $Z_i$ is the Zernike polynomial and $a_i$ is the Zernike coefficient provided by Zemax. Such a method uses some approximations: first, the aberrations are propagated as from a theoretical ray going through the pupil center of each sub-system; secondly, the pupil aberrations are computed independently for each optical sub-system and for each slice. That means we know the chief ray position only for the central slice (in which the PSF center is); for the other adjacent slices, the chief ray position is assumed at the centroid of each part of the cut PSF. In POP, the aberrations are better controlled as the instrument is not decomposed in many optical sub-systems and the wavefront propagation is going directly through the complete optical system, taking the aberrations of all previous pupil planes into account. At the end, the use of POP should be more confident on condition that the aliasing phenomenon is well controlled, in particular for the optical systems with stronger aberrations.

By comparing the two methods which have so different limitations, we can estimate the level of our ``mis-modeling''. Nevertheless, we are more confident in the results obtained with IDL as the SNAP spectrograph is diffraction-limited. We have first compared the PSFs computed from POP and IDL on different optical planes of the spectrograph. No significant difference have been measured for the intermediate optical planes. We have found only a light difference on the detector plane where the PSF image of POP presents a low asymmetry in log scale. This affects slightly the PSF center at the level of a tenth of pixel and we measure a deviation less than 5 $\mu m$ ($\approx$ 1/3 of a pixel) between the PSFs width of POP and IDL. We explain these differences by the aliasing phenomenon that is currently the highest limitations of the POP tool.

We have concluded that the knowledge of the PSF with POP is sufficient to validate the functionality of the demonstrator (alignment, focus). Our final choice to use POP is mainly driven by simplicity. POP is directly linked to the optical design of Zemax; we use only one tool, one language which makes easier the implementation and minimizes the errors. At the end, it goes faster to set-up and to generate a library of PSF. We have to take yet into account the pixel response whose the effects on the under-sampled PSF may be equivalent or larger.

\subsection{Simulation of detector effects}
\label{ss-sec:detector}

\begin{figure}[htbp]
	\centering
	\includegraphics[angle=0,width=0.38\textwidth]{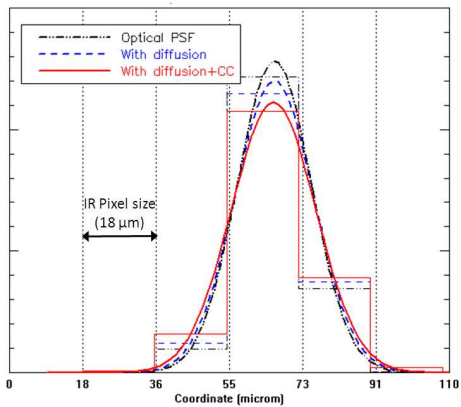} 
	\caption{PSF of the SNAP spectrograph at 1530 nm convolved with the infrared pixel response:  optical PSF (dash-dotted line), optical PSF convolved with the lateral charge diffusion (dotted line) and when we add the capacitive coupling(CC) (solid line). Averaging on the sampling, we find for the FWHM: 1.21 pixels (optical), 1.29 (with diffusion) and 1.35 (with diffusion and CC).}
	\label{fig:psf_pixelresponse}
\end{figure}

To sample on detector pixels, the simulation includes the pixel response such as quantum efficiency, photon noise, readout noise and dark current noise. Furthermore, thanks to the measurements made by the University of Michigan on the infrared detector Rockwell H2RG\#40 used for the demonstrator, we have also taken into account the intra-pixel sensitivity variation, the charge diffusion and the capacitive coupling between pixels. These may have significant effects in particular on the under-sampled PSF. The lateral charge diffusion are simulated with an hyperbolic secant function: $G\left(r\right)=1/{\rm cosh }\left(r/l_d\right)$ where $r$ is the distance from the pixel center. The diffusion length values used is the one given by \cite{Barron2007}: $l_d=1.87 \; \mu m$. The capacitive model suppose equal coupling to each of the four neighbor pixels and negligible coupling to the corner pixels \cite{Brown2006}. The coupling coefficient implemented in the model is of 2\% from  \cite{Barron2007}. Fig.\ref{fig:psf_pixelresponse} shows the PSF computed from the optical simulation and the final PSF convolved with the pixel response. Averaging on the sampling, we estimate the PSF broadening due to the charge diffusion and the capacitive coupling around of 1/6 of a pixel, which is not negligible compared to the required accuracy of 1/10 of a pixel to measure the emission lines (see section \ref{sec:lambda_calib}). We must then take it into account to validate completely the simuation.

\subsection{Simulation of extended sources}

The PSF gives the image on the detector for a monochromatic point-like source. For the demonstrator, we need to simulate the image of spectral and spatial extended sources. These extended sources have been characterized by a brightness profile that describes the intensity distribution in the object plane. The image on the detector of a such source results from the convolution of the PSF with the surface brightness profile. 
To do that, we generate a library of PSF computed at different positions $ \left(x,y\right)$ and wavelength $ \lambda$. Then we can extrapolate the PSF at any entrance point from these library. To be perfectly exact, we would have to extrapolate both the PSF shape and its intensity. Nevertheless, the shape of the demonstrator PSF varies slowly and it is sufficient to take the PSF shape of the library computed at the nearest source point $\left(x,y,\lambda\right)$.  Then we fit only the PSF intensity of the library as a function the point source position and the wavelength using a neural network \cite{Tilquin2006}.
For the PSF library of the demonstrator, we generate 15 000 images covering all the FoV (5 slices) and the wavelength range $\left[0.43-1.7 \; \mu m \right]$. 

\subsection{Conclusion}

To validate completely the simulation, we have compared the PSF modeling with the demonstrator data. We have focused the tests on the infrared range where the pixel response is well known and can be taken into account in the model. Averaging over all the wavelength, we have recovered the PSF width with an  accuracy of 1/6 of a pixel taking into account the lateral charge diffusion and the capacitive coupling. 
The difference is due to a light defocus around $\approx$30 $\mu m$ above the focus specification of 50 $\mu m$. This shows that we have a good understanding of the PSF and of the effects that can degrade the measurements. This comparison with real data gives a good confidence to use the simulation to test and develop procedures for the instrument calibration and for the data reduction.

\section{Wavelength calibration}
\label{sec:lambda_calib}

In this section, we describe the wavelength calibration procedure developed for the demonstrator using extended-like sources. 

\subsection{Objective}

\begin{figure}[htbp]
	\centering
	\includegraphics[angle=0,width=0.45\textwidth]{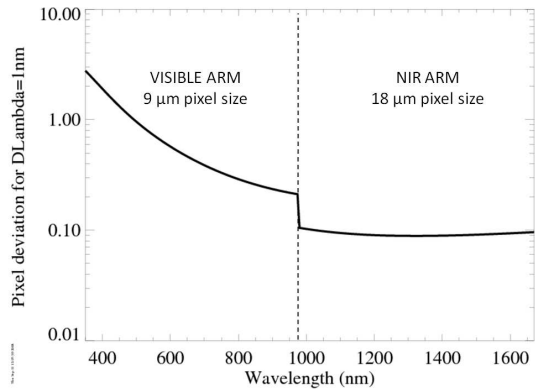} 
	\caption{Ray deviation (in pixel) on the detector associated to a spectral resolution element of one nanometer (scale log).  }
	\label{fig:spec_lambda}
\end{figure}

Because of our configuration, the requirement of 1 nanometer of accuracy corresponds for a large wavelength range to measure the line position at $\approx$1/10 of a pixel.
Figure \ref{fig:spec_lambda} shows the shift of the light (in pixel of the detector) for one nanometer of spectral deviation. In the infrared arm where the spectral resolution is flat, one nanometer corresponds always to a tenth of a pixel. In the visible range, the spectral resolution varies and is higher at short wavelength then one nanometer is covered from two pixels to one tenth of a pixel. The specification of one nanometer is then easier to reach when $\lambda \le 500 $ nm where one nanometer is associated to a shift greater than one pixel. This spectral range will help us to check the wavelength calibration validity. 

\subsection{Data campaigns}

\begin{figure}[htbp]
	\centering 
		\includegraphics[angle=0,width=0.46\textwidth]{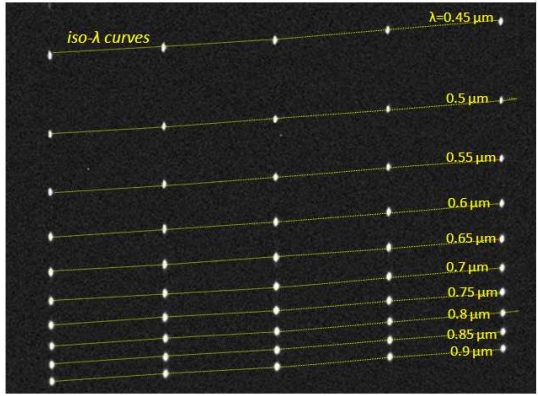}
	\caption{ Detector image of an extended source in the visible range used to fit the demonstrator dispersion curves. In this case, the slice length is lighted only on the central part. On the contrary, the slice width is lighted uniformly.}
	\label{fig:extended_source_visible_dem}
\end{figure}

The two campaigns of the demonstrator have been mainly driven by developing the spectro-photometric calibration adapted to the slicer spectrograph. To do the wavelength calibration, we have used a standard emission line determination procedure.
The emission lines are produced by using an halogen lamp combined with a monochromator. Such an illumination unit provides fine and accurate emission lines (0.35 nm of precision on the wavelength value with a 2 nm width) and prevents us from blended lines more difficult to use in a low spectral resolution configuration. To be independent of the source position in the entrance slit, we use extended spatial sources. As we cannot have directly extended source with the demonstrator, we have ``created'' extended-like sources by taking many images of a same point-like source scanning the FoV. But to minimize the acquisition time, the extended source covers only the central part of the FoV. Fig.\ref{fig:extended_source_visible_dem} shows the resulting image for a set of emission lines in the visible.

The first campaign of the demonstrator at room temperature has provided 23000 CCD images in the visible range. The images processing is classical: we subtract a sky image taking into account both the possible parasite light of the environment and the electronic noise of the CCD. For the second campaign at 140K, we have taken 7000 infrared images using a last generation Rockwell detector HgCdTe. The readout electronics was developed by the IPNL laboratory. We used a Fowler(N) sampling method. Fowler sampling data consists of a reset followed by multiple initial readouts (N frames or one burst), then an exposure, completed by an equal number of final readouts (N). We correct each frame of bias (using a map of reference pixels), quantum efficiency and conversion gain. The final image results of the difference between the mean of the two bursts of N frames. Finally, we correct the brightest pixels using a reference mask of hot pixels. We estimate the error of each pixel as from the Poisson law.

\subsection{Dispersion curves determination}

The emission lines with well known wavelengths are used to fit the dispersion curves.
The spectral dispersion curves describe the relation $\lambda=f_{x_d}\left(y_d\right)$ at a given position $x_d$ between a wavelength at a detector pixel where $x_d$ is the position along the spatial direction (detector rows) and $y_d$ the position along the spectral dispersion direction (detector columns).  

Generally, the spatial distortions (iso-$\lambda$ curves along the spatial direction $x_d$) are fitted to adjust the spectral dispersion curves along the spectrograph entrance slit. As shown in Fig.\ref{fig:extended_source_visible_dem}, for the demonstrator, the spatial distortions are linear for a FoV of five slices (around 150 pixels). We observe also that the spatial distortions inside one slice (imaged on around 20 pixels) are negligible (less than one pixel). We choose then to use an original procedure that does not require to fit the spatial distortions: we choose to fit one spectral dispersion curve for each slice and as the spatial distortions are negligible in one slice, we fit only the spectral dispersion at the slice center. That means we have to determine five dispersion curves for the demonstrator.

 Fitting the spectral lines centers on the detector is easier with extended spatial source since the lines are broaden by the convolution with the slit function.
We named slit function the instrument response after illuminating uniformly the entrance slit with a monochromatic source: namely the ``slit function'' results of the convolution of the slit width with the PSF of the spectrograph. Even if the PSF is sub-sampled (around one pixel at 1 $\mu m$), the slit function spreads it at least over two or three pixels. This helps us to determine the spectral line center on the detector, with a simple barycenter. 

\begin{figure}[htbp]
	\centering
	\begin{tabular}{cc}
	\includegraphics[angle=0,width=0.41\textwidth]{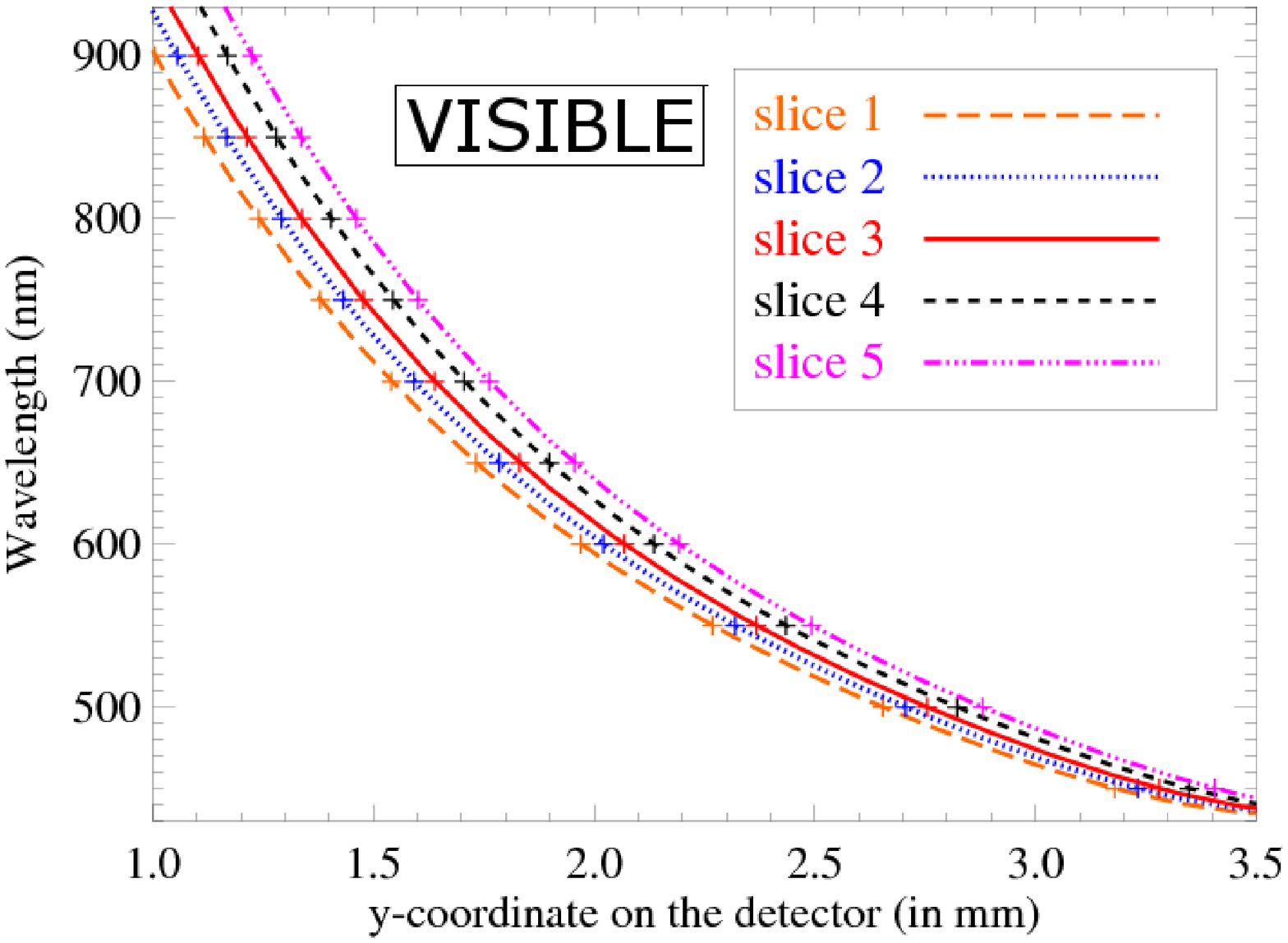} & \includegraphics[angle=0,width=0.41\textwidth]{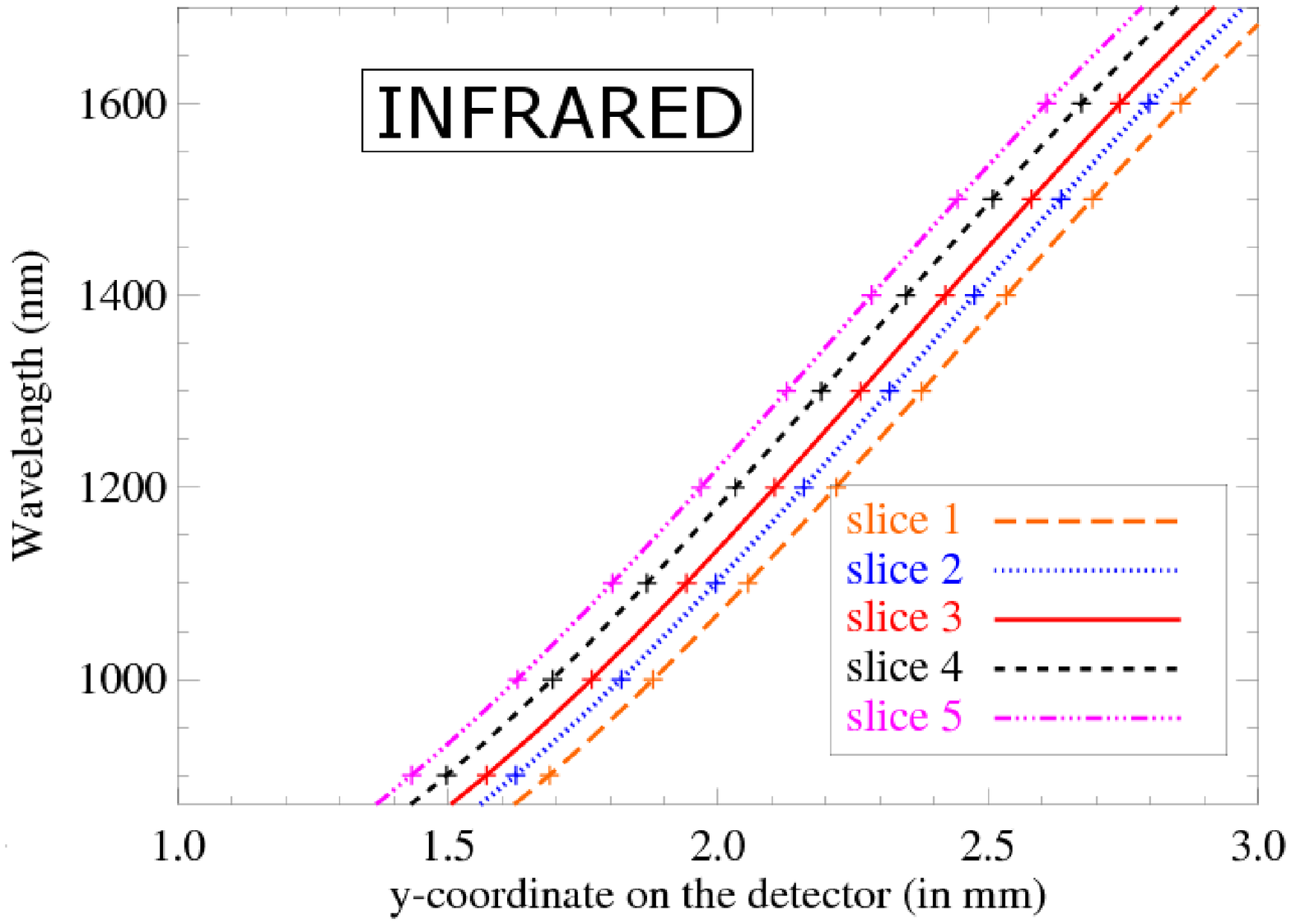} \\
	 	\end{tabular}
	\caption{Spectral dispersion curves fitted from experimental data at the center of each slice, in the visible range (\textit{Left}) and in the infrared range (\textit{Right}). }
	\label{fig:dispersion_curves}
\end{figure}

Fig.\ref{fig:dispersion_curves} shows the experimental dispersion curves fitted at the center of the five slices in the visible and the infrared range. Because of the spectral resolution, the polynomial function is different in the visible and in infrared. In the infrared, where the spectral resolution is almost constant, it is sufficient to use 8 emissions lines separated of about 100 nm in the spectral range [900-1600 nm]. The fitted function for the dispersion curve of the central slice is: $\lambda\left[ \rm {nm}\right]  = f\left( {y_{d} \left[ \rm {mm}\right] } \right) = 1027.38 - 863.13y_{d}  + 641.27 y_{d}^2  -91.36 y_{d}^3  $. In the visible, where the spectral resolution changes, we use 10 emissions lines in the spectral range [450-900 nm] by step of 50 nm and we use a polynomial function of five orders. The adjusted dispersion curve for the central slice in the visible is: $\lambda\left[ \rm {nm}\right]  = f\left( {y_{d} \left[ \rm {mm}\right] } \right) = 1688.7 - 8.62y_{d}  + 0.016y_{d}^2  + 3.7 \cdot 10^{ - 5} y_{d}^3  - 1.83 \cdot 10^{ - 7} y_{d}^4  + 1.94 \cdot 10^{ - 10} y_{d}^5 $.  

The experimental curves have been compared with the simulation and are been found to be in a complete agreement at 95\% confidence level. This validates both the simulation and the experimental measurements. 

\section{Wavelength measurement of a point-like source}
\label{sec:slit_effect}

To test the accuracy of the calibration at the nanometer level, the dispersion curves have been used to measure the wavelength of a point-like source. 
Point-like sources, in contrary of extended-like sources, are sensitive to the positioning inside the slit. We have then tested the ability to recover a line position without an ``a priori'' knowledge  of the source position  in the visible and the infrared. 

An image slicer spectrograph has the same properties as a classical slit-spectrograph.  In particular, the off-centering of a point-like source inside the entrance slit can create a shift of the line center on the detector.  This is classically called the ``slit effect'' and can lead to an error on the measured  wavelength.  
As we already emphasize, the SNAP spectrograph configuration can degrade the line determination since the spectrograph has a low spectral resolution ($\approx 100$) and the spectrograph is optimized with a low-sampled PSF for the visible range and under-sampled in the infrared range.  The emission line fitting is then more sensitive to the detector sampling and to the position inside the FoV.
We have not particularly addressed the case of extended-like sources as it is a direct extension of the sources used previously to set the dispersion curves, and as they are less affected by the slit effect since the entrance slit is lighted uniformly or in a symmetrical way. Then this effect should be evaluated and corrected only for point like sources.
 
\begin{figure}[htbp]
	\centering
	\begin{tabular}{cc}
	\includegraphics[angle=0,width=0.48\textwidth]{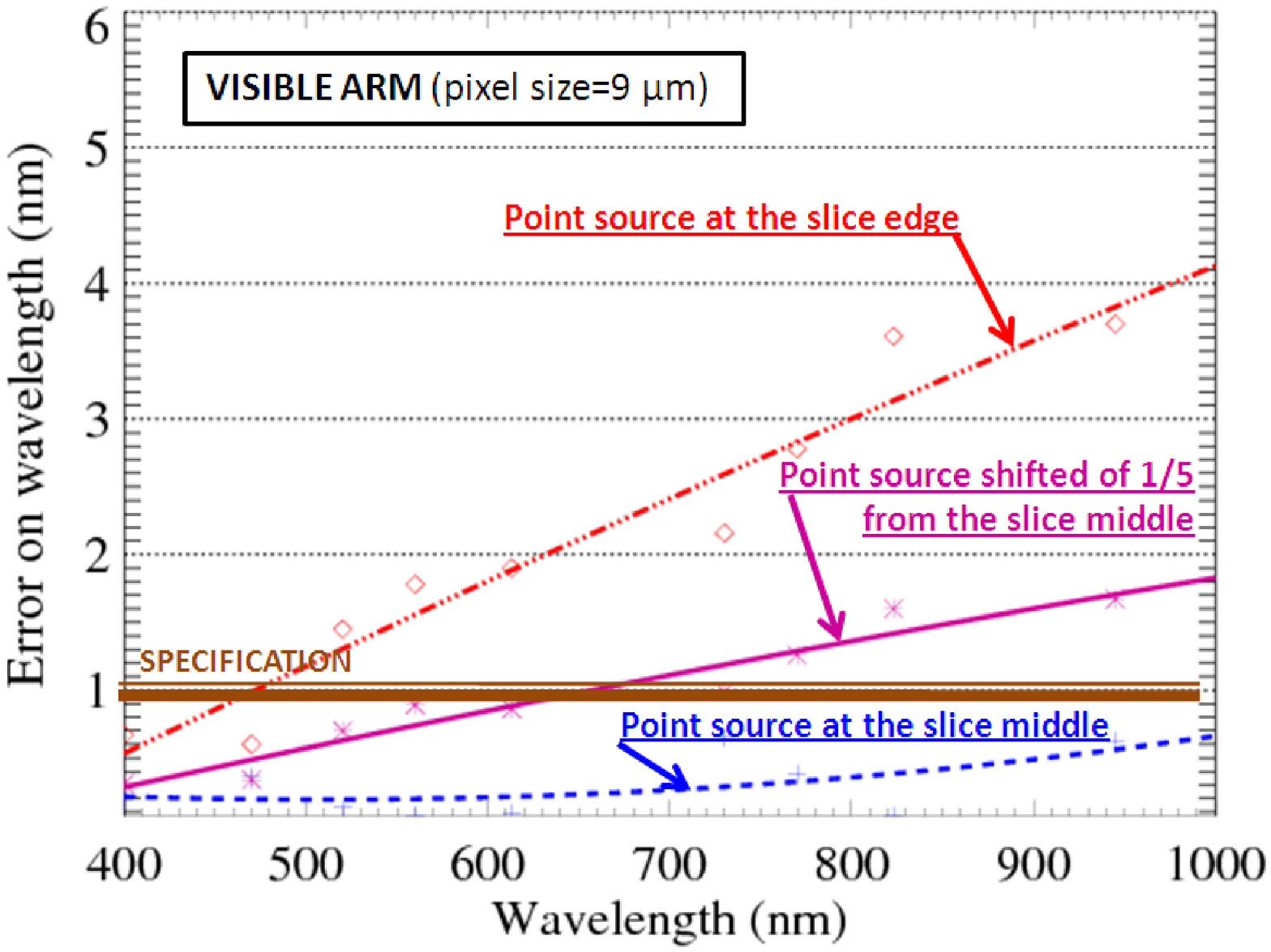} & \includegraphics[angle=0,width=0.48\textwidth]{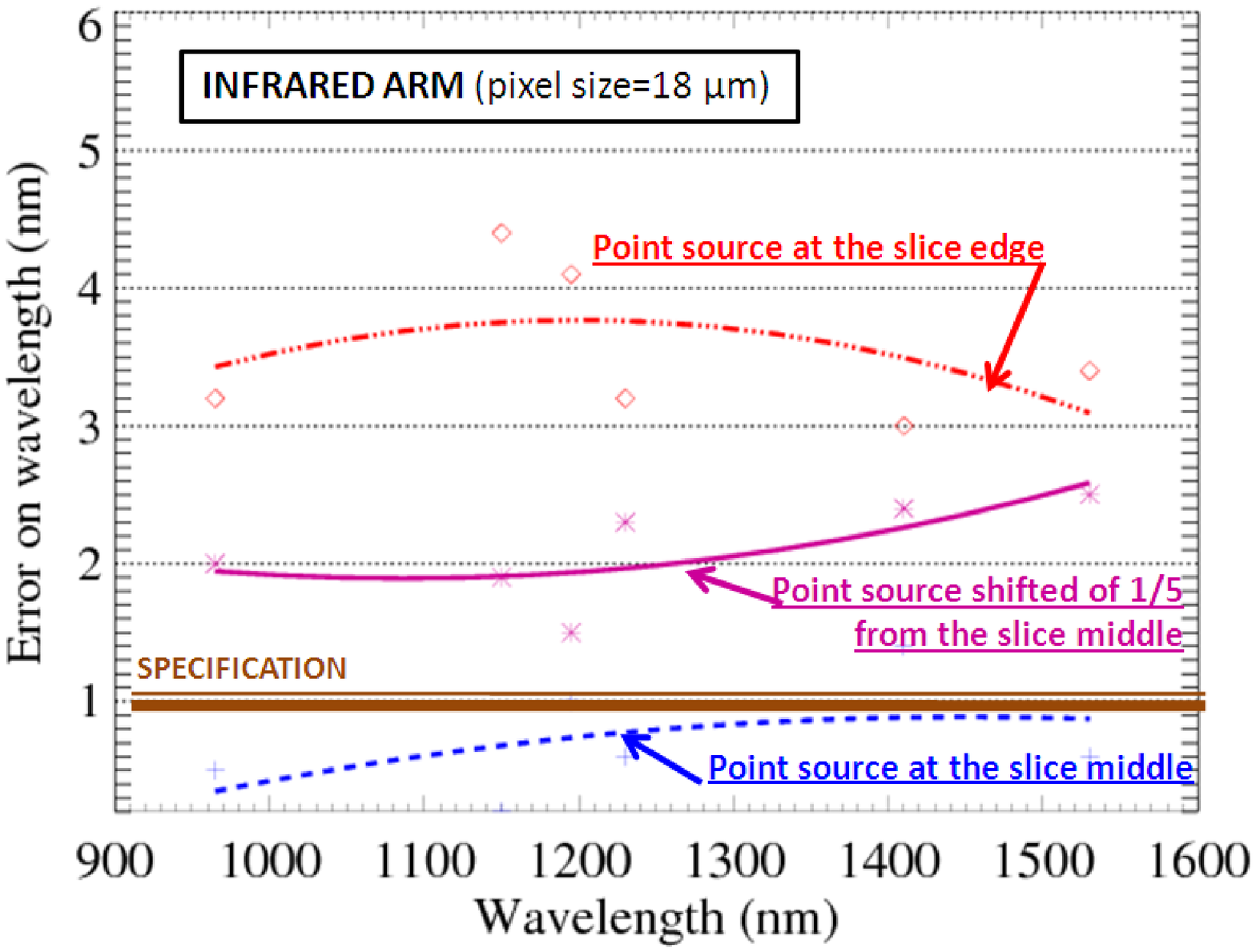} \\
	 	\end{tabular}
	\caption{Error on the wavelength measurement due to slit effect (simulation) as a function of the wavelength in the visible \textit{(Left)} and infrared \textit{(Right)} arm. Three positions of point-like source are represented: at the middle, shifted of a quarter of slice and on the slice edge.  }
	\label{fig:slit_effect_uv_ir}
\end{figure}

We have first estimated the slit effect using the simulation. We represent on Fig.\ref{fig:slit_effect_uv_ir} the deviation of the measured wavelength compared to the true one in nanometer, when we move the point source inside the slice (e.g in the slit):  on the center of the slice, we recover the position within the specification (1 nm) but, close to the slice edges, we observe large discrepancies (up to 4 nanometers). 
In the infrared, as it was shown on Fig.\ref{fig:spec_lambda}, the spectral resolution is constant and one nanometer is lighted on 1/10 of pixel. When changing the position inside the slit, the PSF is cut differently and the position can be shifted up to 4 nm. In the visible, where the spectral resolution varies, the deviation depends more of the wavelength and increases when the resolution decreases. In this case, the error depends of the PSF size and of the position. Only low wavelength values where the resolution is high and the PSF is small are inside the specification. In other cases, the resolution is decreasing faster than the PSF increases, which translates to an evolving slit effect.

It is then clear than to be able to find back the good wavelength values, we need to correct or control this effect for all wavelengths. We have then developed a procedure that can recover a good accuracy thanks to the slicer properties.

\subsection{Method}

In a slit-spectrograph, the only way to correct the slit effect is to known accurately the position inside the slit.  Nevertheless, in a spatial environment, we would like to relax on the requirement of the telescope pointing system. With a slicer, we can use the information of each slice to correct the slit effect and improve significantly the line determination without knowing the position of the object.

As the PSF spreads out up to 5 slices, we can find back the spectrum of the point-like source up to 5 slices on the detector. Instead of using one spectrum in the central slice, we can use then the spectra imaged on different slices. The advantage is that the sampling of the spectrum on each slice is not the same in both the spatial and the spectral directions thanks to the spatial distortions. Averaging the spectra on 5 adjacent slices acts like a ``spatial or spectral dithering".
Furthermore, thanks to the slicer `cut-out'', the PSF intensity distribution from one slice to another will vary with the point source position. This helps to recover the relative position of the point source onto the slicer plane by looking at the flux distribution between the slice. The result of the averaging is that we recover the position without knowing the light source position inside the slit and so correct the slit effect without any new procedure or mechanism.

\section{Results}

We present now the measurement of ``monochromatic'' emission lines sources in the visible and the infrared range.
To quantify the accuracy, we compute the difference between the initial wavelength $\lambda_i$ and the extracted wavelength $\lambda_e$ of the observed emission lines: $\Delta \lambda=\lambda_i-\lambda_e$. This absolute error $\Delta \lambda$ is expressed in nanometer. This error includes both the error due to the accuracy of the dispersion curves and the error due to the accuracy of the lines center determination on the detector.

Practically, for $N$ slices $\left(N \leq 5\right)$, we use a barycenter to extract the line center on the detector and we deduce the associated wavelength $\left\{\lambda_i\right\}_{1\leq i \leq N}$ from the spectral dispersion curves. Then we deduce the final wavelength by averaging the $N$ wavelengths $\left\{\lambda_i\right\}_{1\leq i \leq N}$ weighted by the flux measured in each slice. We estimate first the error $\Delta \lambda$ for a classical slit-spectrograph method, considering only the spectrum in one slice (central slice). Then we take the average of the wavelength extracted of each slice as explained previously.

\begin{figure}[th]
	\centering
	\begin{tabular}{|cc|}
	\hline
 \includegraphics[angle=0,width=0.48\textwidth]{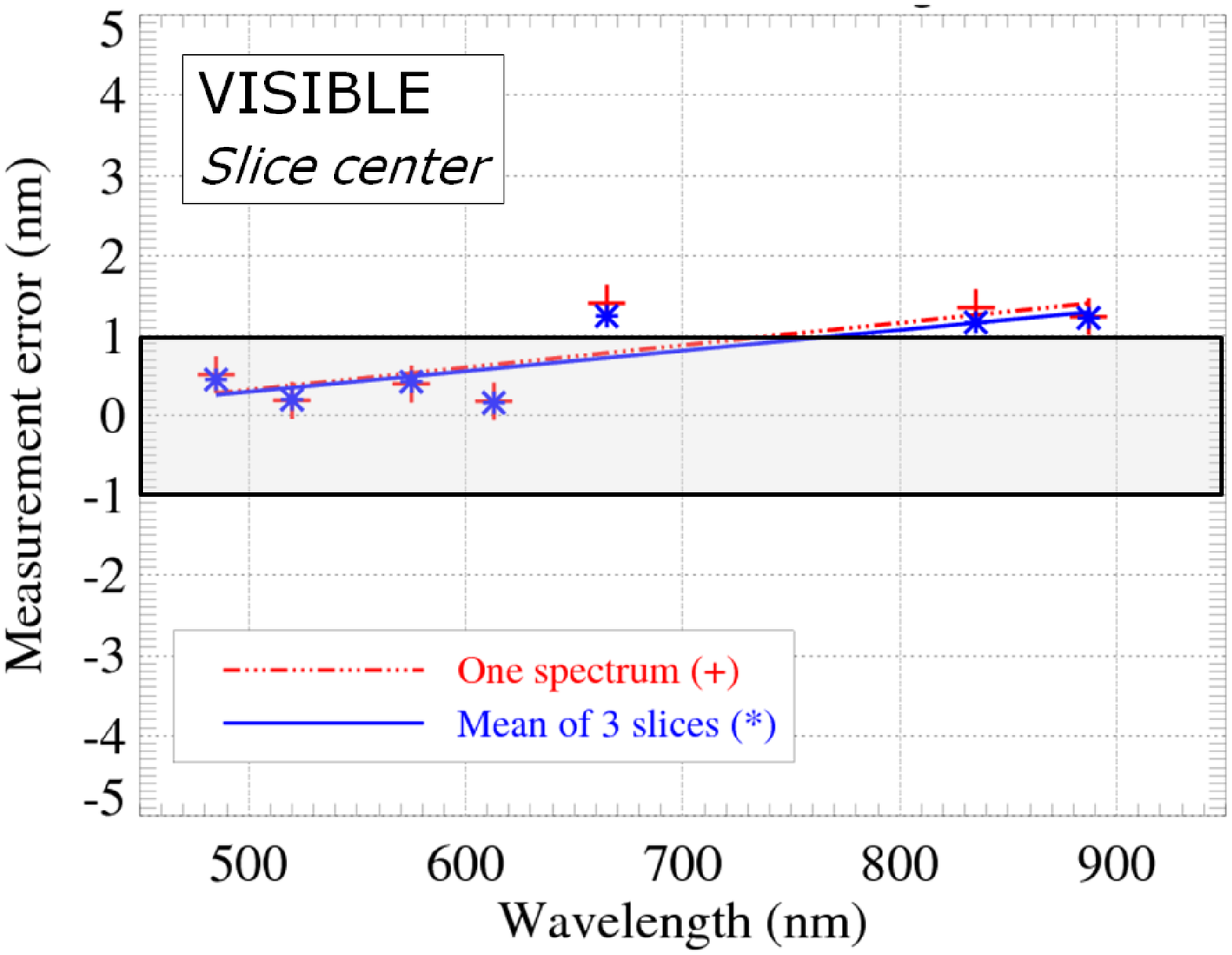} & \includegraphics[angle=0,width=0.48\textwidth]{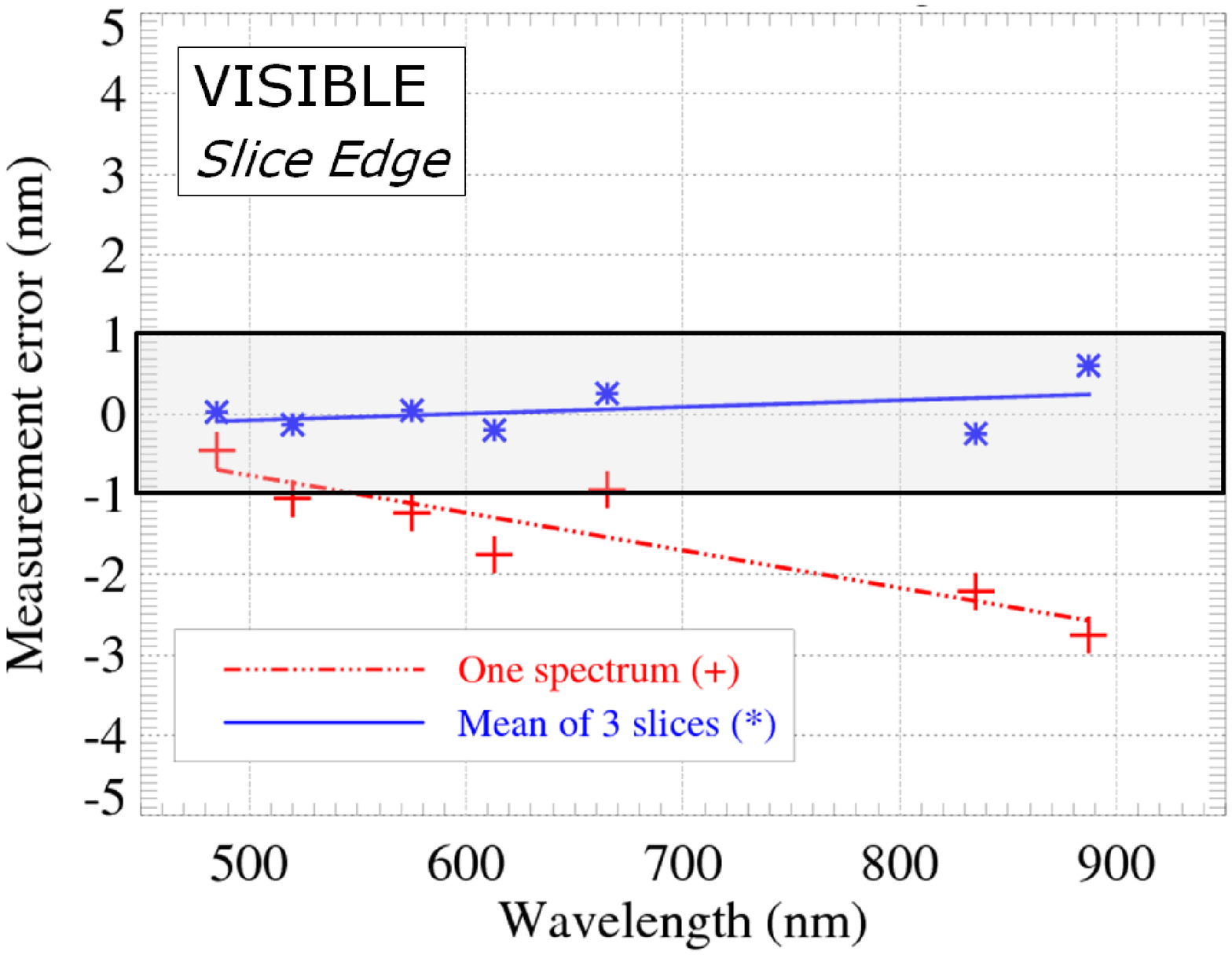} \\
	\hline
	 	\end{tabular}
	\caption{Experimental error (in nanometer) as a function of the wavelength in the visible  when the point-like source is at the slice center (left) and at one slice edge (right). Using only the spectrum in the central slice (dash-dotted lines), the error on the wavelength is sensitive to the slit effect, especially at the slice edge. Averaging the spectra on 3 slices improves the wavelength accuracy better than 1 nm.}
	\label{fig:calib_lambda_result1}
\end{figure}

\begin{figure}[bh]
	\centering
	\begin{tabular}{|cc|}
	\hline
	 \includegraphics[angle=0,width=0.48\textwidth]{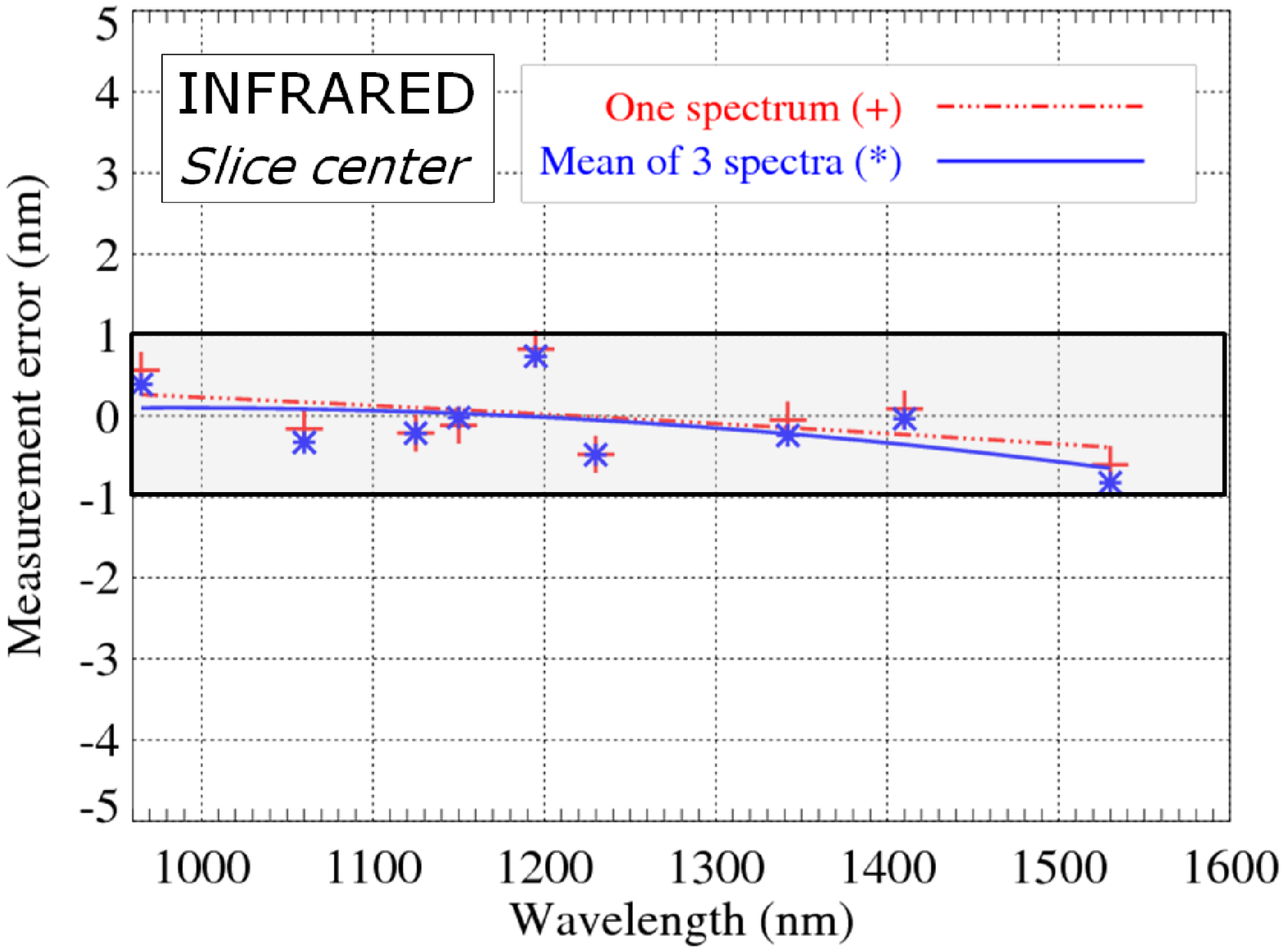} &
	 	 \includegraphics[angle=0,width=0.48\textwidth]{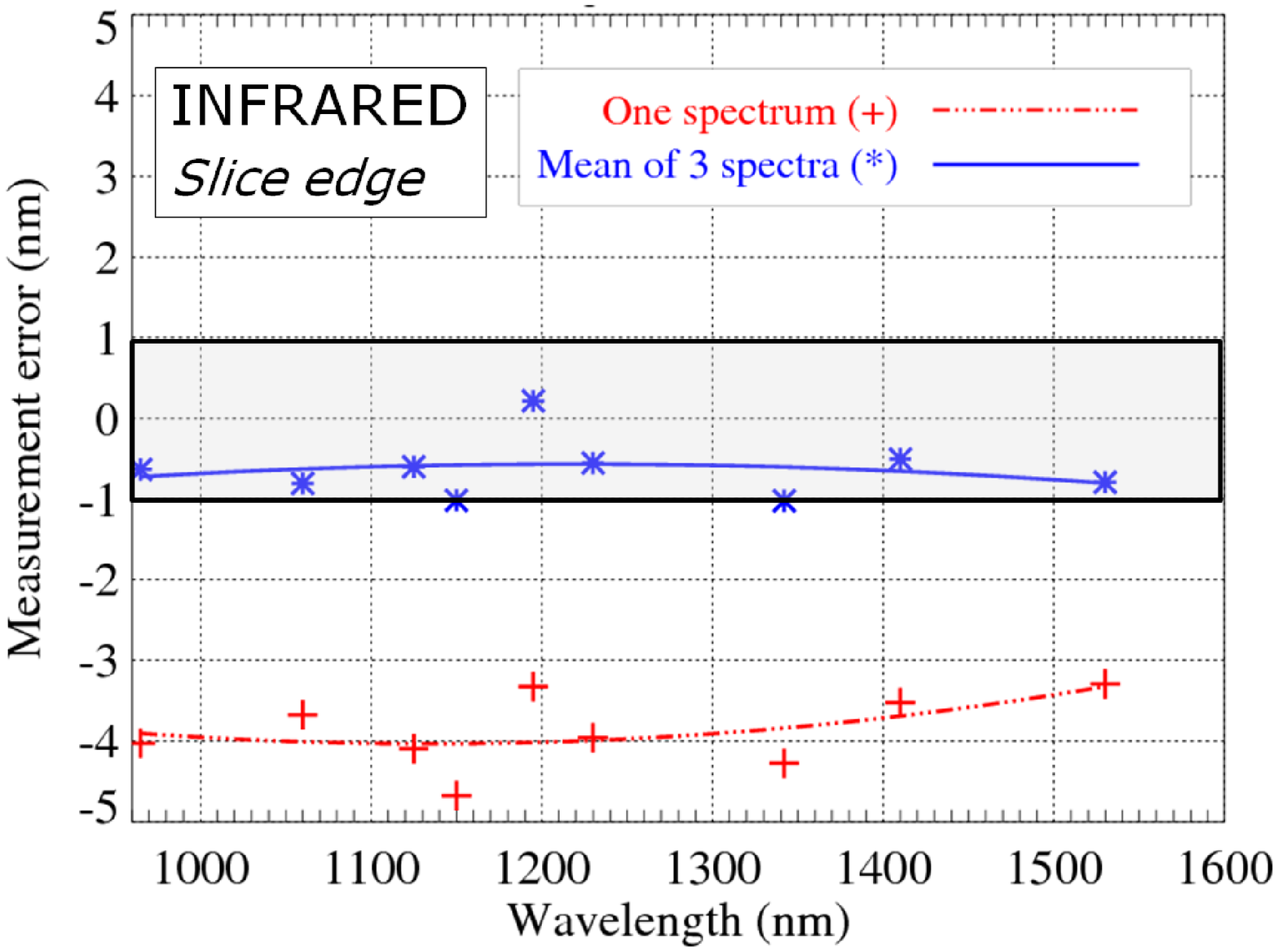}\\
	\hline
	 	\end{tabular}
	\caption{Experimental error (in nanometer) as a function of the wavelength in the infrared when the point-like source is at the slice center (left) and at one slice edge (right).}
	\label{fig:calib_lambda_result2}	
\end{figure}

Fig.\ref{fig:calib_lambda_result1} and  Fig.\ref{fig:calib_lambda_result2} shows the result as a function of the wavelength  at two positions in the slice (slice center and edges) both for the visible and infrared range. At the slice center, averaging the spectra on 3 slices has no impact on the error since we do not need to correct the slit effect. At contrary, on the edge, the value is shifted as expected by the slit effect and the averaging of the spectra on 3 slices allows to recenter in the specification.
In the visible, we see that the slit effect increases with the wavelength, due to the spectral resolution and it should be corrected. In the infrared, as the spectral resolution is constant, the error due to the slit effect is constant over all the range (about 4 nm) and is also corrected better than one nanometer. We have observed that the averaging of the spectra on 5 slices does not improve the accuracy on the wavelength. At contrary, in the infrared range, the mean of spectra on 5 slices brings more error than the level of improvement since the most distant slices are dominated by the noise.

\begin{figure}[th]
	\centering
	\begin{tabular}{|cc|}
\hline
	 \includegraphics[angle=0,width=0.48\textwidth]{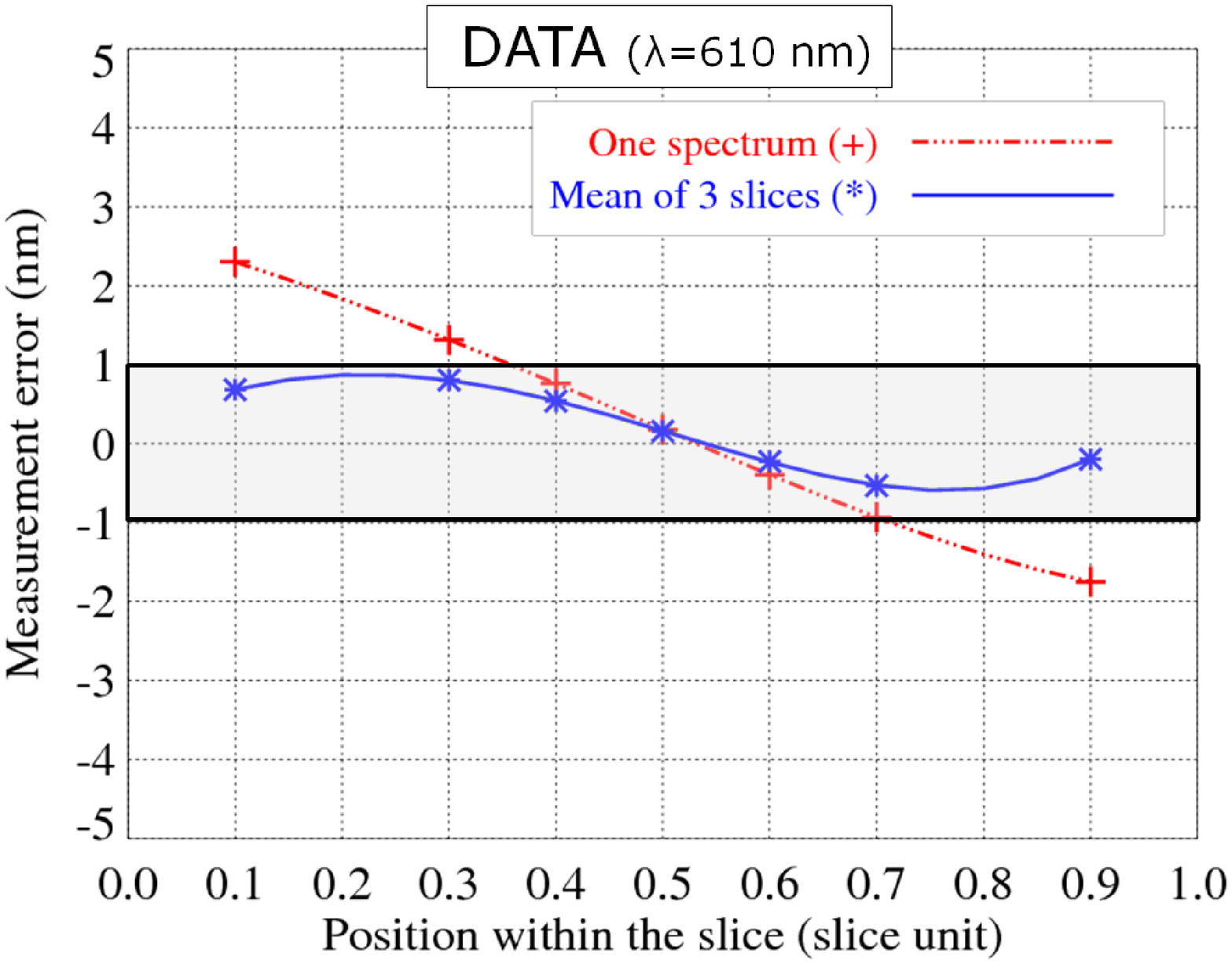} &
	 \includegraphics[angle=0,width=0.48\textwidth]{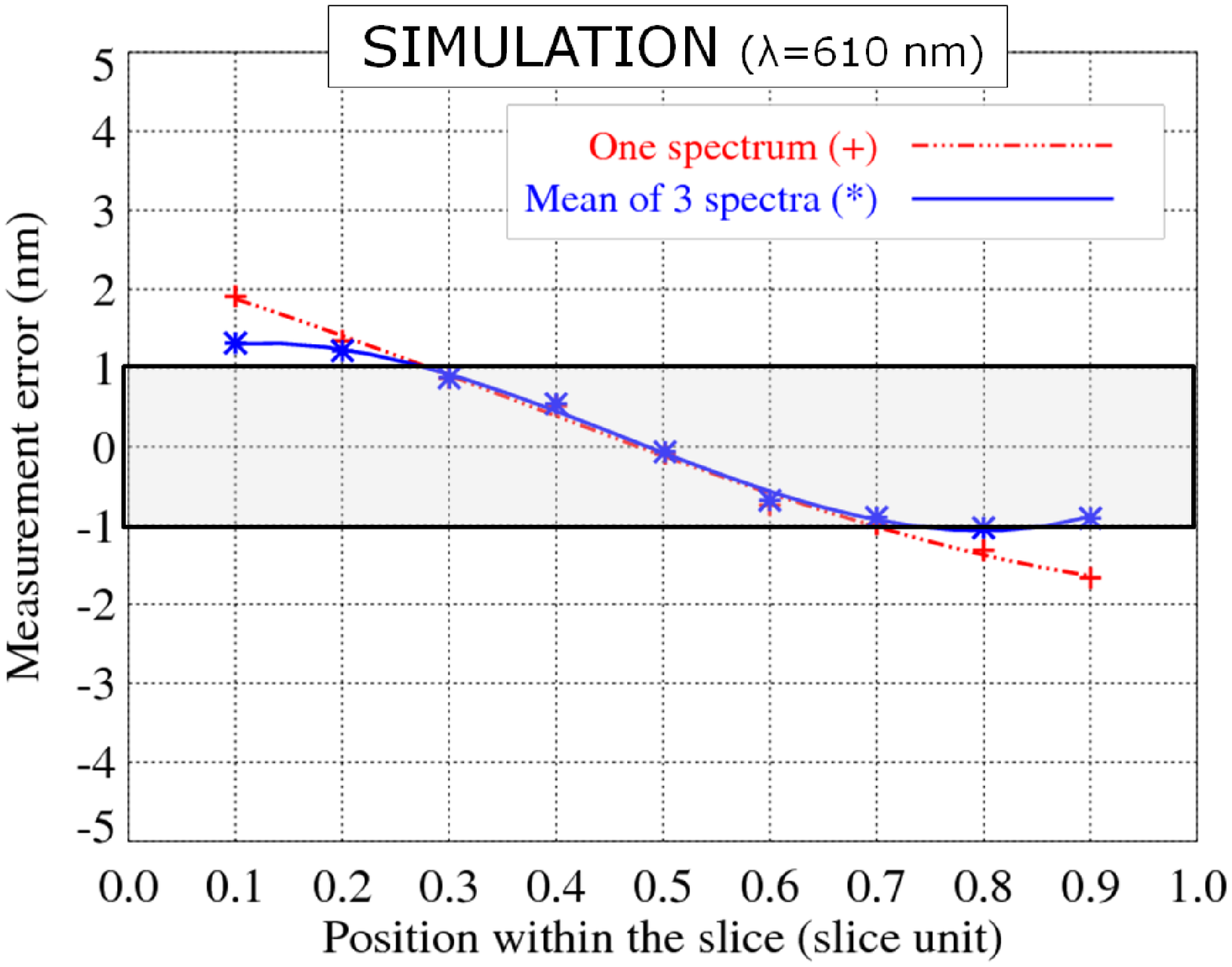}\\
	\hline
	 	\end{tabular}
	\caption{Simulated and experimental error (in nanometer) on the wavelength measurement of one emission line at 610 nm  as a function of the point-like source position.}
	\label{fig:calib_lambda_wrty_result1}
\end{figure}

\begin{figure}[bh]
	\centering
	\begin{tabular}{|cc|}
\hline
	\includegraphics[angle=0,width=0.48\textwidth]{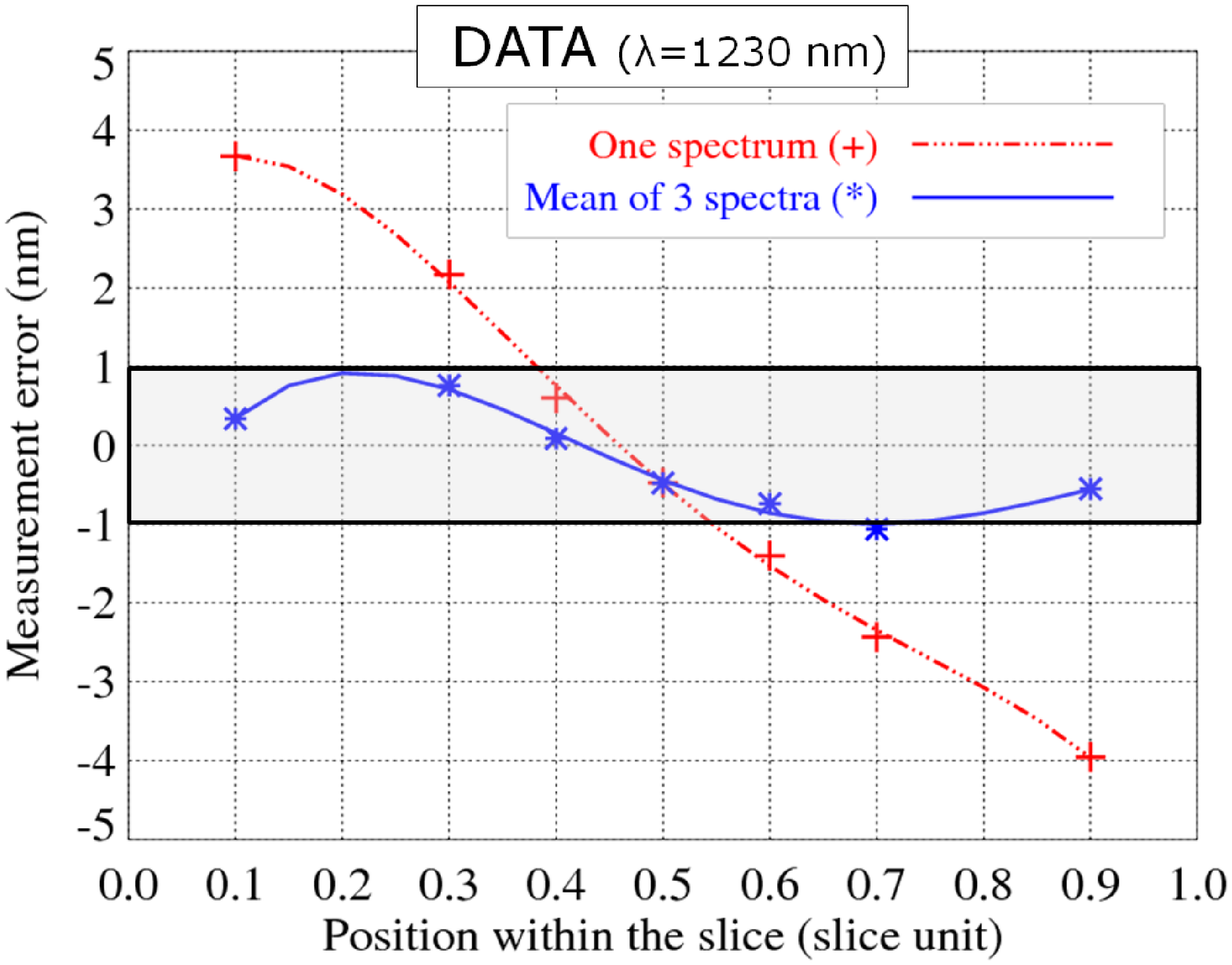} &
	 \includegraphics[angle=0,width=0.48\textwidth]{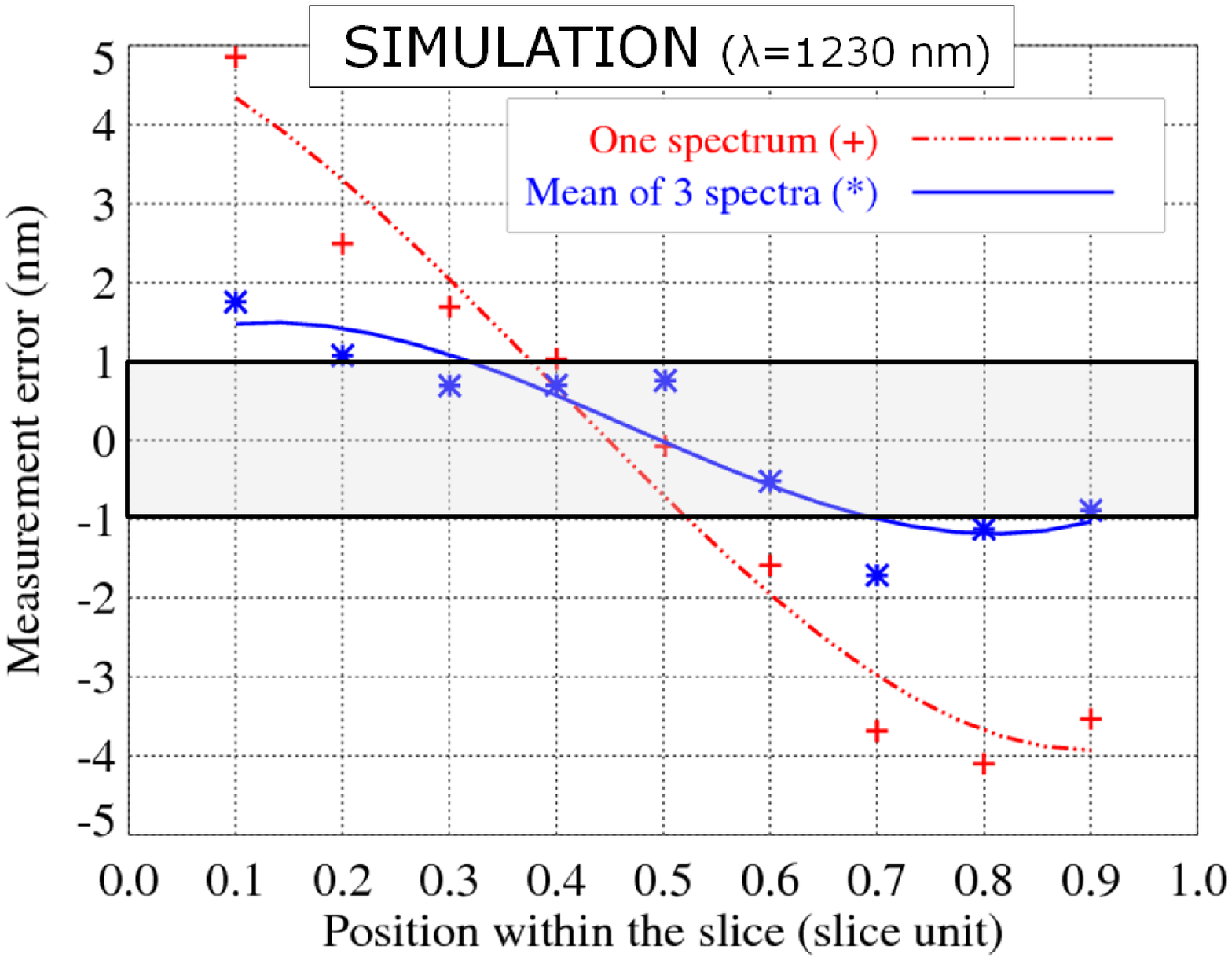}\\
	\hline
	 	\end{tabular}
	\caption{Simulated and experimental error (in nanometer) on the wavelength measurement of one emission line at 1230 nm as a function of the point-like source position.}
	\label{fig:calib_lambda_wrty_result2}
\end{figure}

On Fig.\ref{fig:calib_lambda_wrty_result1} and Fig.\ref{fig:calib_lambda_wrty_result2}, we represent the error as a function of the slice position and compare it with the simulation both in the visible at 610 nm and in the infrared at 1230 nm. We observe the same behavior at the center and edge of the slice: the simulation is in good agreement. Surprisingly, the error on the wavelength seems better with the demonstrator. We explain this deviation by the asymmetry of the simulated PSF which is responsible of at least 1/10 of pixel shift on the PSF center (as we have detailed in paragraph \ref{ss_sec:psf_model}). This phenomenon can explain the observed fluctuations with the simulation.

\clearpage

\begin{figure}[htbp]
	\centering
	\includegraphics[angle=90,width=0.5\textwidth]{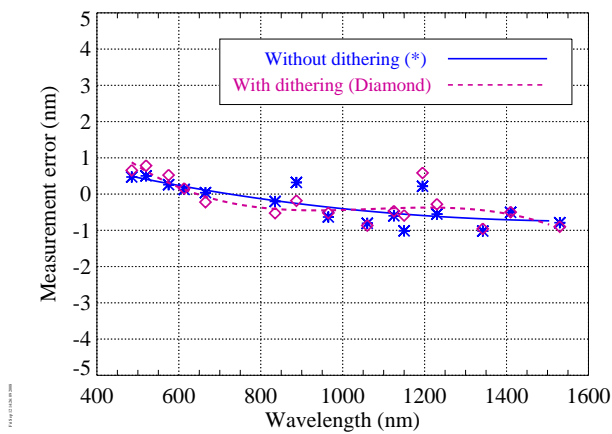} 
	\caption{Measurement error (in nanometer) as a function of the wavelength when we average only the spectra on three slices (solid line) and when we use both the mean of the spectra on 3 slices and the spatial dithering technique (dashed line).}
	\label{fig:result_lambda_dithering}
\end{figure}

Finally, we have also tested the ``spatial dithering" technique. We simulate the spatial dithering by combining four images of one point-like source shifted in the FoV. The shift is in accordance with the jitter telescope(random shift with a Gaussian distribution of 0.03 arc second RMS\footnote{Root Mean Square}). The wavelength is computed from the average of the extracted wavelength of the four images. Fig.\ref{fig:result_lambda_dithering} shows the measurement error obtained with the demonstrator as a function of the wavelength when we apply the spatial dithering technique or not, compared to the result of the three slices averaging. We observe no improvement on the measured error. We concluded that averaging the spectra on 3 slices is sufficient to achieve the expected accuracy since averaging the 3 spectra acts also as a spatial dithering.

\section{Conclusions}

With the demonstrator, we have been able to prove than we can measure wavelengths at a precision of one nanometer in the visible and infrared range within the SNAP specifications. We have proved that it is possible to correct the slit effect thanks to the slicer properties, using the fact that each slice provides one spectrum of the source on the detector. The averaging of the spectra on 3 slices is equivalent to a dither as the sampling of the 3 spectra is different in both the spectral and spatial distortions. This has been further confirmed by applying a real spatial dithering: we see that there is no improvement on combining 4 images at 4 positions in the FoV. The averaging on 3 slices compensates also the low spectral resolution and allows to measure the wavelength of emission lines with an accuracy better than one nanometer even in sub-sampling condition.

The procedure not required to known either the position of the source, either to extract it precisely inside a slit. This means that, thanks to the slicer, we can also relax the requirement of the telescope pointing system. In a slit spectrograph, the telescope should be pointed a 0.01 arc second, to have a 1/10 pixel precision. In a slicer, we can relax this requirement without necessarily degrading the accuracy on the measurement wavelength. This simplifies the calibration procedure without adding any procedure or mechanism.

This last result allows to be confident on the extraction of emission lines in a sub-sampling configuration when we use a slicer. As the galaxy are extended source, we expect to measure the redshift from emission line at a precision of $ \approx 0.003 \times \left(1+z\right)$. Thanks to the slice average method, the application to the point-like source as the calibration of a star or for a supernovae will be simplified. In particular, with an accuracy of one nanometer, we expect to be able to measure the SiII absorption feature of supernovae Ia at a precision of $\approx 400 \; km/s$ on all the redshift range. 

\acknowledgments

\section*{Acknowledgments}

We thank the entire demonstrator team from LAM (Laboratoire d'Astrophysique de Marseille), CPPM (Centre de Physique des Particules de Marseille) and IPNL (Institut de Physique Nucélaire de Lyon) for the pleasure to have prepared this work with them. We thanks the SNAP collaboration for their support and help. We thank in particular M.Lampton for discussions and useful comments. This work was supported by CNRS (Centre National de la Recherche Francaise), CNES (Centre National d'Etudes Spatiales) and LBNL (Lawrence Berkeley Laboratory). M-H. Aumeunier is supported by the spatial french agency CNES.

\end{document}